\pdfoutput=1
\documentclass[
reprint,
superscriptaddress,
showpacs,preprintnumbers,
 amsmath,amssymb,
 aps,
]{revtex4-1}

\usepackage{graphicx}
\usepackage{dcolumn}
\usepackage{bm}
\usepackage{hyperref}

\begin{document}

\preprint{arXiv:xxxx.2014}

\title{The detection of back-to-back proton pairs in Charged-Current neutrino interactions \\
with the ArgoNeuT detector in the NuMI low energy beam line}

\author{R.~Acciarri}
\affiliation{Fermi National Accelerator Laboratory, Batavia, IL 60510 USA}
\author{C.~Adams}
\affiliation{Yale University, New Haven, CT 06520 USA}
\author{J.~Asaadi}
\affiliation{Syracuse University, Syracuse, NY 13244 USA}
 \author{B.~Baller}
 \affiliation{Fermi National Accelerator Laboratory, Batavia, IL 60510 USA}
 \author{T.Bolton}
 \affiliation{Kansas State University, Manhattan, KS 66506 USA}
 \author{C.~Bromberg}
 \affiliation{Michigan State University, East Lansing, MI 48824 USA}
 \author{F.~Cavanna}
  \affiliation{Yale University, New Haven, CT 06520 USA}
  \affiliation{Universit\`a dell'Aquila e INFN, L'Aquila, Italy}
  \author{E.~Church}
  \affiliation{Yale University, New Haven, CT 06520 USA}
 \author{D.~Edmunds}
 \affiliation{Michigan State University, East Lansing, MI 48824 USA}
 \author{A.~Ereditato}
 \affiliation{University of Bern, Bern, Switzerland}
 \author{S.~Farooq}
 \affiliation{Kansas State University, Manhattan, KS 66506 USA}
 \author{B.~Fleming}
\affiliation{Yale University, New Haven, CT 06520 USA}
 \author{H.~Greenlee}
\affiliation{Fermi National Accelerator Laboratory, Batavia, IL 60510 USA}
 \author{G.~Horton-Smith}
 \affiliation{Kansas State University, Manhattan, KS 66506 USA}
 \author{C.~James}
\affiliation{Fermi National Accelerator Laboratory, Batavia, IL 60510 USA}
 \author{E.~Klein}
\affiliation{Yale University, New Haven, CT 06520 USA}
 \author{K.~Lang}
 \affiliation{The University of Texas at Austin, Austin, TX 78712 USA}
 \author{P.~Laurens}
 \affiliation{Michigan State University, East Lansing, MI 48824 USA}
 \author{R.~Mehdiyev}
 \affiliation{The University of Texas at Austin, Austin, TX 78712 USA}
 \author{B.~Page}
 \affiliation{Michigan State University, East Lansing, MI 48824 USA}
 \author{O.~Palamara}
 \thanks{Corresponding author: ornella.palamara@yale.edu}%
\affiliation{Yale University, New Haven, CT 06520 USA}
\affiliation{INFN - Laboratori Nazionali del Gran Sasso, Assergi, Italy}
 \author{K.~Partyka}
\affiliation{Yale University, New Haven, CT 06520 USA}
 \author{G.~Rameika}
\affiliation{Fermi National Accelerator Laboratory, Batavia, IL 60510 USA}
 \author{B.~Rebel}
 \affiliation{Fermi National Accelerator Laboratory, Batavia, IL 60510 USA}
 \author{M.~Soderberg}
 \affiliation{Syracuse University, Syracuse, NY 13244 USA}
 \affiliation{Fermi National Accelerator Laboratory, Batavia, IL 60510 USA}
 \author{J.~Spitz}
\affiliation{Yale University, New Haven, CT 06520 USA}
 \author{A.M.~Szelc}
\affiliation{Yale University, New Haven, CT 06520 USA}
 \author{M.~Weber}
 \affiliation{University of Bern, Bern, Switzerland}
\author{T.~Yang}
\affiliation{Fermi National Accelerator Laboratory, Batavia, IL 60510 USA}
\author{G.P.~Zeller} 
\affiliation{Fermi National Accelerator Laboratory, Batavia, IL 60510 USA} 

\collaboration{ArgoNeuT Collaboration}

\date{\today}
             
\begin{abstract}
Short range nucleon-nucleon correlations in nuclei ({\it NN} SRC) carry important information on nuclear structure and dynamics. {\it NN} SRC 
have been extensively probed through two-nucleon knock-out reactions in both pion and electron scattering experiments.
We report here on the detection of two-nucleon knock-out events from neutrino interactions and discuss their topological features
as possibly involving {\it NN} SRC content in the target argon nuclei.
The ArgoNeuT detector in the Main Injector neutrino beam at Fermilab has recorded a sample of 30  fully reconstructed charged current events where the leading muon is accompanied by a pair of protons at the interaction vertex, 19 of which have both protons above the Fermi momentum of the Ar nucleus. Out of these 19 events, four 
are found with the two protons in a strictly back-to-back high momenta configuration directly observed in the final state and can be
associated to nucleon Resonance pionless mechanisms involving a pre-existing short range correlated {\it np} pair in the nucleus. Another fraction (four events) of the remaining 15 events have a reconstructed back-to-back configuration of a {\it np} pair in the initial state, a signature compatible with one-body Quasi Elastic interaction on a neutron in a SRC pair.
The detection of these two subsamples of the collected  ($\mu^-+2p$) events suggests that mechanisms directly involving nucleon-nucleon SRC pairs in the nucleus are active and can be efficiently explored in neutrino-argon interactions with the LAr TPC technology.
\end{abstract}

\maketitle
{\it Introduction} - The systematic study of the impact of nuclear effects on the determination of neutrino cross sections in the ``few-GeV region" and on neutrino oscillation parameters,  has developed into a very active field of theoretical and experimental research over the last decade \cite{NuInt}. 
Effects of long-range two-nucleon processes in the nuclear target 
are widely recognized as a source of apparent cross section enhancement in 
 the neutrino Charged Current Quasi-Elastic (CC QE) response \cite{morfin}, sufficient to provide a possible 
 solution to the excess of neutrino event rate in the energy range around $\sim$1 GeV observed by MiniBooNE
 \cite{MiniBooNE}. 
 Since emitted nucleons in Cherenkov detectors are typically below detection threshold, the so-called CCQE-like sample can in fact find contributions from both single nucleon knock-out events from a genuine CC QE reaction and 
from  two-nucleon emission events produced for example by  two-body meson-exchange currents (MEC).\\
\indent However, different nuclear processes other than those involving two-nucleon currents can also lead to two-nucleon ejection: 
 initial state short-range nucleon-nucleon correlations ({\it NN} SRC) and final state interactions (FSI). 
The realization of consistent models including all these nuclear effect is now being actively pursued 
(e.g. see \cite{martini,benhar,amaro,bodek,sobczyk,nieves,giusti,gran,benhar2} and references therein) 
as well as their implementation in MonteCarlo generators (MC).
Direct experimental investigations on the nature of nuclear effects and their impact on the predicted rates, final states, and kinematics of  neutrino QE interactions
are even more compelling.
 The availability of new experimental techniques such as the Liquid Argon Time Projection Chamber (LArTPC), with its full 3D imaging, precise calorimetric energy reconstruction and efficient particle identification, is opening new perspectives for detailed reconstruction of final state event topologies from neutrino-nucleus interactions.
A different approach and methodology in neutrino data analysis with LArTPCs is now being developed, based on the topological categorization of event samples collected by the ArgoNeuT experiment (Argon Neutrino Test) in the ``few-GeV" energy range - a region of particular interest to future long-baseline neutrino oscillation experiments. \\
In this Letter we present and briefly discuss the observation of a sample of events detected in ArgoNeuT during the exposure to the
NuMI LE (Neutrinos at the Main Injector, Low Energy option) beam at Fermi National Accelerator Laboratory (FNAL).
The specific final state topology which we have focused on is a pair of energetic protons at the interaction vertex accompanying the leading muon. This topology may provide hints for {\it NN} SRC in the target nucleus when the protons of the pair appear with high-momentum (exceeding the Fermi momentum) and 
in strong angular correlation. In particular, in analogy with findings from electron- and hadron-scattering experiments,
a neutrino CC QE interaction on 
a neutron in a SRC pair is expected to produce back-to-back protons in the CM frame of the interaction,
whereas a CC pionless resonance reactions (CC RES) involving a SRC pair may produce back-to-back protons in the Lab frame.\\
\indent {\it The ArgoNeuT experiment} - The ArgoNeuT detector \cite{ArgoNeuT} is a  $47\times 40\times 90~$cm$^{3}$ active volume LArTPC ($\sim 240$ kg of LAr) with the longer dimension along the beam direction. ArgoNeuT collected several thousands of  $\nu_\mu$ and $\bar\nu_\mu$ charged current interactions during an extended run in 2009-10 at FNAL.
In particular, {\it neutrino} events were acquired from both configurations of the NuMI LE beam, namely from a $\sim2$ week run (8.5$\times$10$^{18}$ proton on target, POT) in $\nu$-beam mode and a $\sim5$ month run (1.25$\times$10$^{20}$ POT) in anti-$\nu$-beam mode where neutrinos are a significant component of the flux.  Average $\nu$ energy in $\nu$-beam mode is $\langle E_\nu\rangle\simeq 4$ GeV, 
while in anti-$\nu$-beam mode neutrinos are at higher energies, $\langle E_\nu\rangle\simeq10$ GeV. 
The beam is along the $\hat {\rm z}$ axis of our Lab reference frame and the drift direction of the TPC is along the horizontal $\hat {\rm x}$ axis.
The read-out configuration of the two instrumented wire-planes and the individual wire signal recording allow for the 3D imaging of the ionization tracks in the LArTPC volume as well as the calorimetric reconstruction of the 
energy deposited by individual charged particles along their ionization trail.
Tracks are reconstructed in 3D by combining associated 2D clusters that are identified in both views of the TPC, 
having the drift coordinate in common.
The energy loss along the track is estimated in steps, and the total
energy deposited along the track is obtained by summing over the steps.
For particles slowing down and stopping inside the LArTPC active volume (contained tracks), the energy loss
as a function of distance from the end of the track is used as a powerful method for particle identification. 
For non-contained muons escaping in the forward direction, the MINOS Near Detector, a 0.98 kton magnetized steel-scintillator calorimeter (MINOS-ND) \cite{MINOS-ND}
located just downstream from ArgoNeuT, provides charge sign and momentum reconstruction. 
Track matching between ArgoNeuT and MINOS-ND is based on a common timestamp from the accelerator complex and track alignment requirements \cite{ArgoNeuT}. \\
\indent {\it Event reconstruction} - 
Taking advantage of the reconstruction capabilities of LArTPCs, individual events are categorized in terms of exclusive topologies observed in the final state and used to explore the evidence of nuclear effects in neutrino-argon interactions.\\
Muon neutrino CC events are inclusively selected by requiring a negatively charged muon in MINOS-ND matching a track originating from an interacting vertex in the ArgoNeuT detector. From this sample, the main class of pionless exclusive CC topologies ($\mu^-$+N{\it p}), usually labelled as ``CC 0-pion", is then extracted \cite{Partyka}. 
In this class of events the leading muon can be accompanied by any number (N=0, 1, 2, 3, $\ge 4$) of protons in final state. 
The reconstruction of  the individual proton kinematics (kinetic energy and 3-momentum) is determined with good angular resolution and down to a low proton kinetic energy threshold of 21 MeV. Details of the reconstruction procedure can be found in \cite{ArgoNeuT}. 
In principle, neutrons can  also be emitted in these events, however ArgoNeuT has a very limited capability to detect neutrons emerging from the interaction vertex. This is because the detector size is too small to have significant chances to allow neutrons to convert into visible protons in the LArTPC volume before escaping.
 
 In this Letter we search for possible hints of nucleon-nucleon correlations in the ArgoNeuT data, by specifically looking for exclusive $\nu_\mu$ CC 0-pion events with N=2 protons in the final state, i.e. the ($\mu^-+2 p$) triple coincidence topology.
This data sample amounts to  30 events in total -- 19 collected from anti-$\nu$-beam mode and 11 in $\nu$-beam mode. Both proton tracks are required to be fully contained inside the fiducial volume of the TPC and above the energy threshold [we note that events where one of the two protons is below threshold fall in the ($\mu^-+1p$) sample].
From detector simulation, the overall acceptance for the ($\mu^-+2p$) sample is estimated to be around 35\%, dominated by the 
 requirement  of containment in the fiducial volume. 
 According to a GENIE \cite{GENIE} MC simulation, in either beam modes about 40\% of the ($\mu^-+2p$) events are due to CC QE interactions and about  40\% to CC RES pionless interactions.
For the anti-$\nu$-beam mode ($\nu$-beam mode) run the efficiency corrected $\frac{(\mu^-+2{\it p})}{(\mu^-+{\rm N}{\it p})}$ event ratio is 21\% (26\%) and the $\frac{(\mu^-+2{\it p})}{{\rm CC-inclusive}}$ ratio is 2\% (4\%).

\begin{figure}
\vspace{-0.3cm}
  \includegraphics[height=.26\textheight]{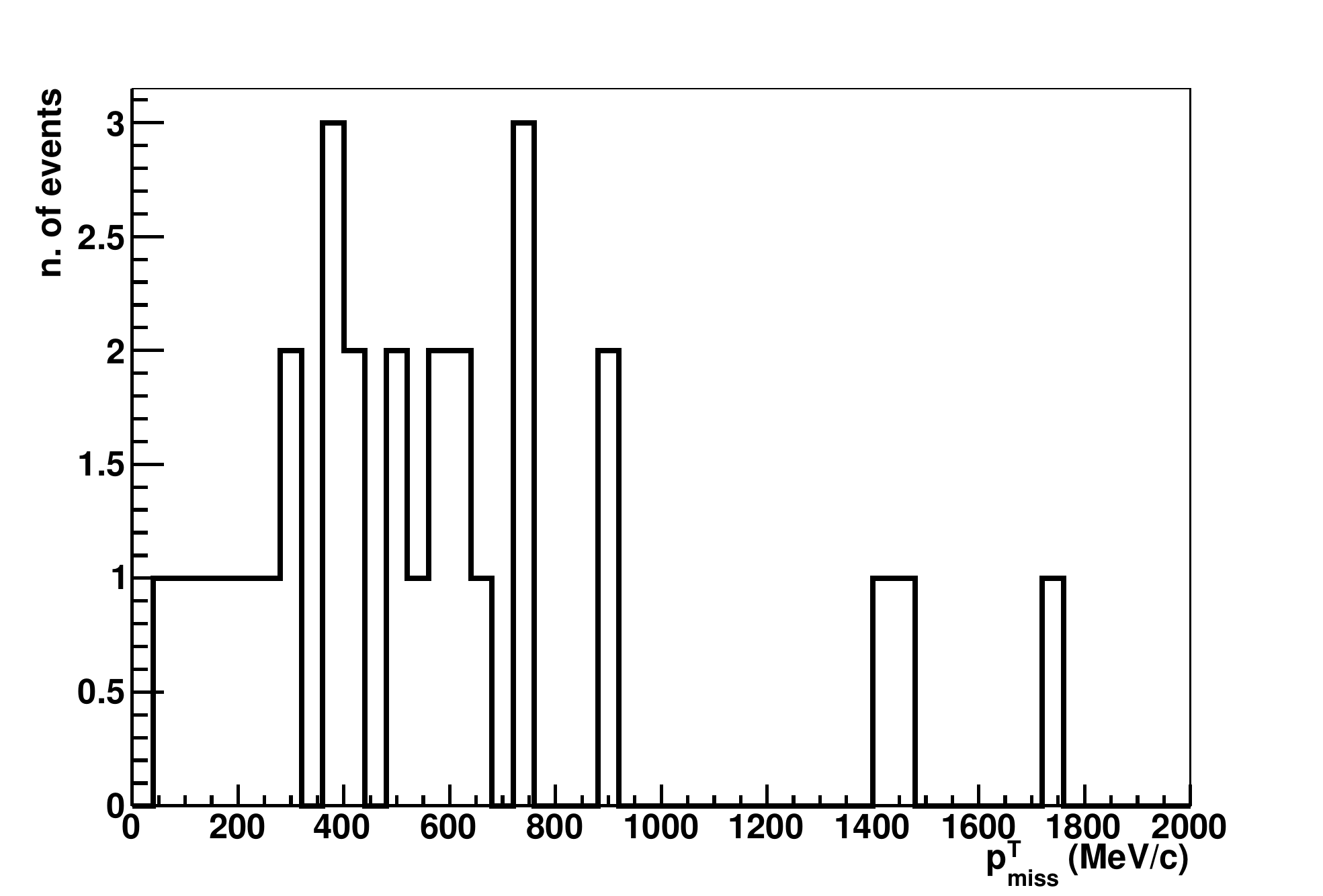}
  \caption{\label{transverse_momentum}{Missing transverse momentum distribution for the 30 events in the ($\mu^-+2p$) sample.}}
\end{figure}

 The $\nu_\mu$ CC reaction leading to the observation of the ($\mu^-+2p$) event sample undergoes the general 
 4-momentum conservation law:
\begin{equation}
k_\nu + P^i_{\rm A} = k_\mu + p_{p1} + p_{p2} + P^f_{\rm X}
\label{eq:4mom-cons}
\end{equation}
where $k_\nu$ and $k_\mu$ refer to the initial neutrino and final muon 4-momenta, $p_{p1}{\rm~and~}p_{p2}$ refer to the two protons in the final state.  $P^f_{\rm X}$ refers to the recoiling nuclear system {\rm X} and $P^i_{\rm A}$ to the target Ar nucleus.
If we assume the two ejected protons originate from an initial state correlated pair, the 
initial pair configuration is neutron-proton $np$, as required by charge conservation in the CC reaction, and $P^i_{\rm A}=P^i_{\rm A-2}+P^i_{np}$.\\
Experimentally measurable observables are the 3-momentum of the muon, determined from the matched track in ArgoNeuT and MINOS-ND, the sign of the muon provided by MINOS-ND, and the energy and direction of propagation of the two protons measured by ArgoNeuT. 
The target nucleus (A=Ar) is at rest in the Lab and the CM of the correlated {\it np} pair is assumed to be (nearly) at rest in it. 
\begin{figure}
 \vspace{-0.3cm}
  \includegraphics[height=.27\textheight]{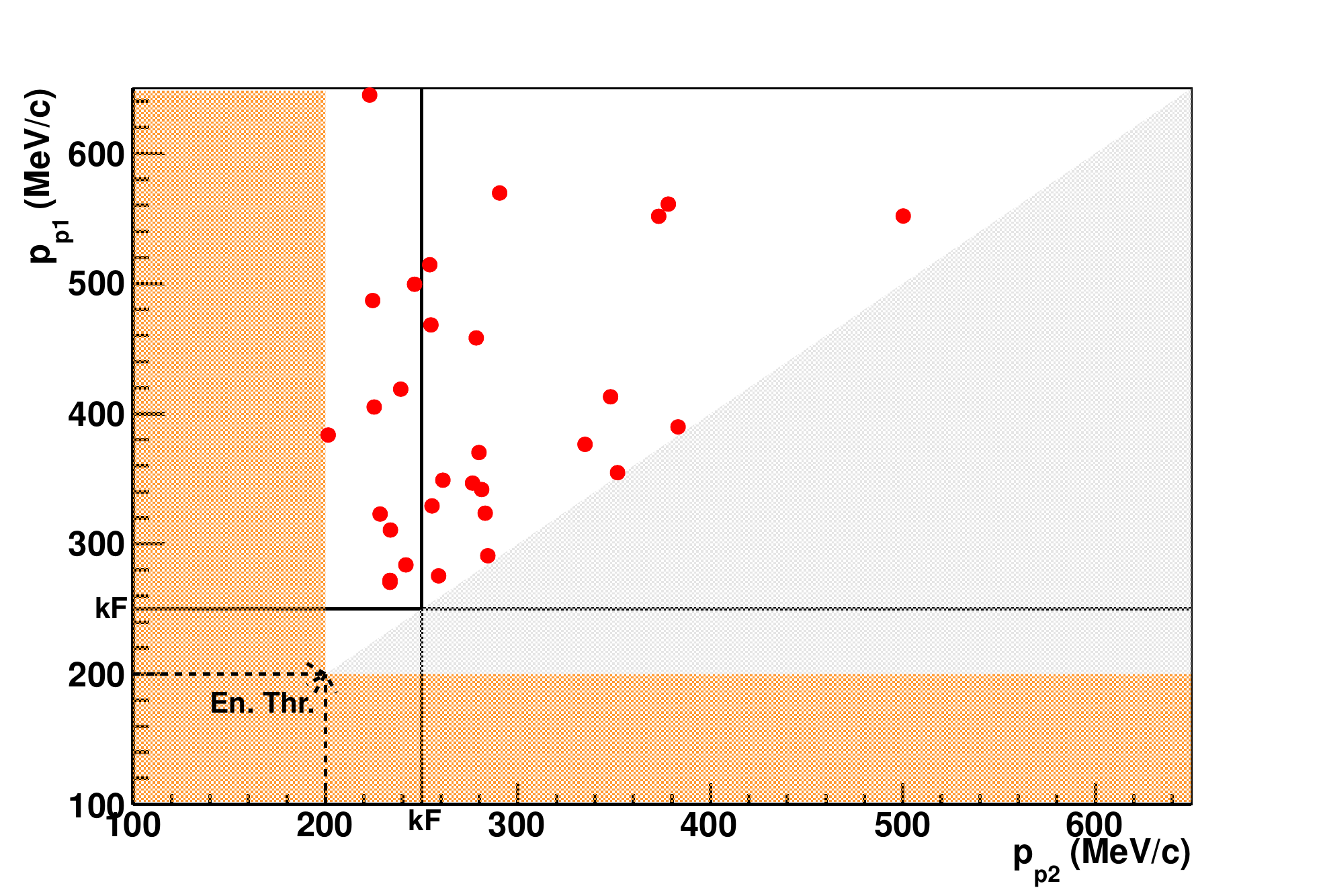}
  \caption{\label{phigh_vs_plow}{Momentum, p$_{p1}$, of the most energetic proton in the pair  vs. momentum, p$_{p2}$, of the other (least energetic) proton for the 30 events in the ($\mu^-+2p$) sample. The Fermi momentum in argon (line) and the momentum corresponding to the detection threshold in ArgoNeuT (dashed) are also indicated.}}
\end{figure}
\begin{figure}
\vspace{-0.3cm}
  \includegraphics[height=.27\textheight]{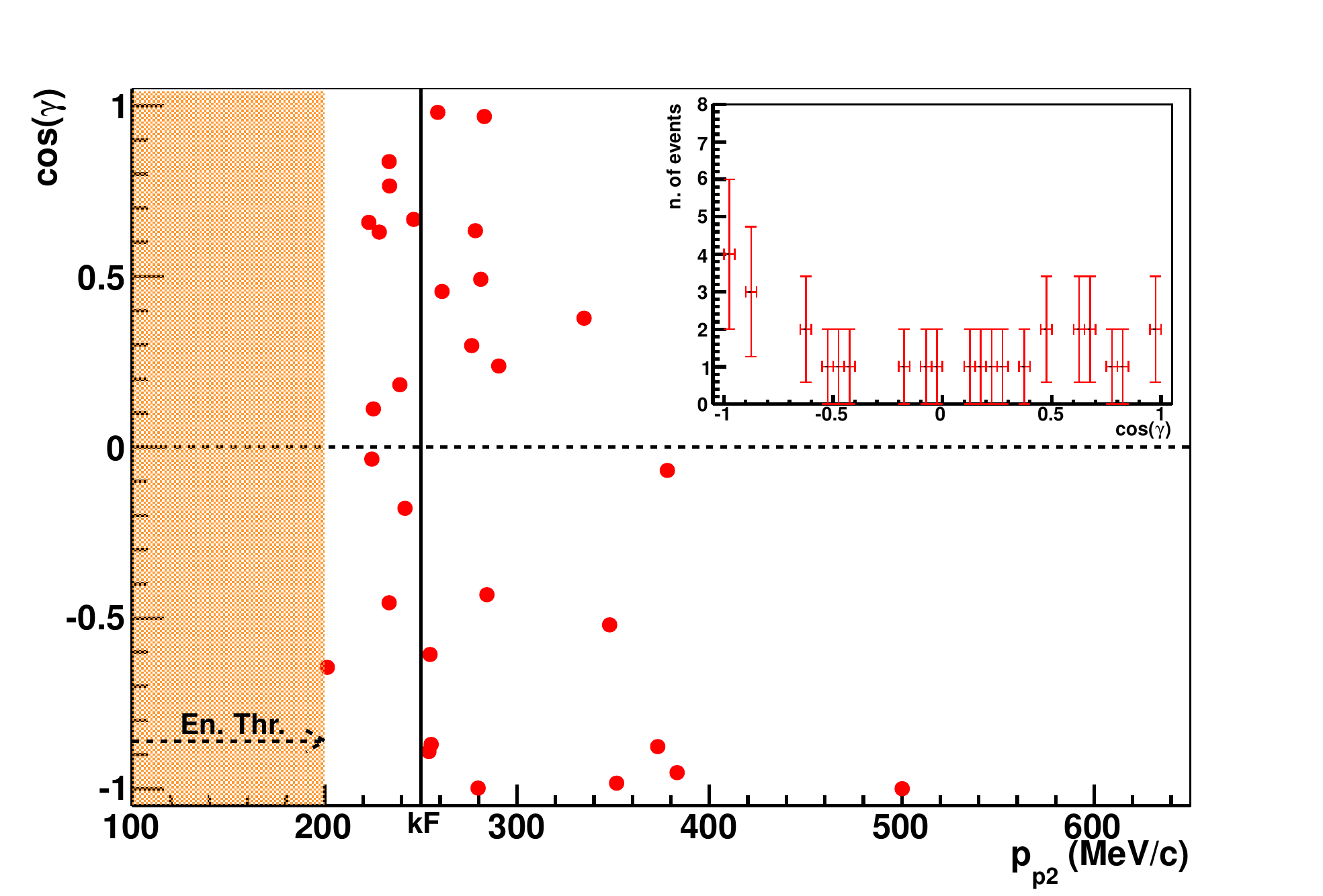}
  \caption{\label{cosgamma_vs_mom}{Cosine of the angle $\gamma$  between the two protons (Lab frame) vs. the momentum of the least energetic proton in the pair for the 30 events in the ($\mu^-+2p$) sample. In the inset is the distribution of cos($\gamma$).}}
\end{figure}
The nuclear system X in final state, an excited (A-2)$^*$ bound state or any other unbound state, is undetected 
and we take its momentum components equal to the momentum components of the missing 4-momentum vector $P_{\rm miss}$. 
The direction of the incident neutrino is along the $\hat{\rm z}$ axis, therefore 
the missing transverse momentum (in the $\hat{\rm x},\hat{\rm y}$ plane) is directly measurable
as ${\rm P}^{\rm T}_{\rm miss}=-({\rm k}^{\rm T}_\mu+{\rm p}^{\rm T}_{p1}+{\rm p}^{\rm T}_{p2})$ from Eq.(\ref{eq:4mom-cons}).
This corresponds to the transverse momentum of the residual nuclear system ${\rm P}^{\rm T}_{\rm A-2}$.
The missing energy component $E_{\rm miss}$ is here defined as the energy expended to remove the nucleon pair from the nucleus.\\
The final state proton momenta determined from the energy measurement of fully contained tracks are reported in Fig.\ref{phigh_vs_plow},
with the scatter plot of the higher vs the lower momentum of the $pp$ pair in the ($\mu^-+2p$) sample. 
Most of the events (19 out of 30) have both protons above the Fermi momentum of the Ar nucleus (k$_F\simeq$250 MeV \cite{Moniz}, solid lines in Fig.\ref{phigh_vs_plow} - we take here an average value for the proton and the neutron Fermi momentum).\\
The angle in space $\gamma$ between the two detected proton tracks at the interaction vertex
is directly measured in the Lab reference frame.
The scatter plot of Fig.~\ref{cosgamma_vs_mom} shows the cosine of the $\gamma$ angle vs. the momentum of the least energetic proton in the pair. The cos($\gamma$) distribution is also reported (inset of Fig.~\ref{cosgamma_vs_mom}). It is interesting to note that four of the nineteen 2$p$-events above the Fermi momentum are found with the pair in a back-to-back configuration (cos($\gamma$)$<$-0.95). \\
The missing transverse momentum measured from  the unbalanced momentum of the triple coincidence ($\mu^-+2p$) in the plane transverse  to the incident neutrino direction is shown in Fig.~\ref{transverse_momentum}. The tail at very high ${\rm P}^{\rm T}_{\rm miss}$ can be explained as due to events with undetected energetic neutron(s) emission.

The incident energy is not confined to a single value but distributed in a broad
$\nu$-beam energy spectrum. 
From energy conservation in Eq.(\ref{eq:4mom-cons}), the incident neutrino energy for the ($\mu^-+2p$) events is given by 
$E_\nu=(E_\mu+T_{p1}+T_{p2}+T_{\rm A-2}+E_{\rm miss})$.
An estimate can be inferred from the final state particles (muon and two protons) measured kinematics.
The last two terms are small corrections: 
the residual nuclear system is undetectable, 
however a lower bound for its recoil kinetic energy 
can be calculated using the measured transverse missing momentum as $T_{\rm A-2}\approx({\rm P}^{\rm T}_{\rm miss})^2/2 M_{\rm A-2}$.
The missing energy includes two terms, namely the two nucleon separation energy for argon \cite{sep-en} 
and the actual excitation level of the residual nucleus.
We set its total value to a constant $E_{\rm miss}$=30 MeV. This is an approximation of the average energy to 
remove a $np$ pair from a Ar nucleus extrapolated from single nucleon removal energy spectra for Ar nuclei \cite{Tyren}.\\
From the reconstructed neutrino energy and the measured muon kinematics,
the components of the 4-momentum transfer ($\omega$,$\vec{\rm q}$) can eventually be inferred.\\
The muon momentum resolution is 5-10\% \cite{ArgoNeuT}. The proton angular resolution (1-1.5$^\circ$, depending on the track length) and the proton energy resolution (about 6\% for protons above the Fermi momentum) are estimated by MC simulation. The overall resolution in our neutrino energy and transfer momentum reconstruction is dominated by muon momentum resolution, as in CC interactions the muon takes the largest fraction on the incident neutrino energy.
\indent {\it Discussion} - Nucleon-nucleon correlations are essential components of modern potentials describing 
the mutual interaction of nucleons in nuclei.
 The strong, repulsive short-range correlations ({\it NN} SRC) cause
the nucleons to be promoted to states above the Fermi level in the high-momentum tail of the nucleon momentum distribution \cite{Pandharipande}.
Thus, SRC cause nucleons to form pairs with large relative momentum and small center-of-mass momentum, i.e. pairs of nucleons with large, back-to-back momenta. Due to {\it NN} tensor correlations,  SRC pairs are dominantly in iso-singlet (deuteron like) state $(np)_{I=0}$
\cite{schiavilla}.\\
Two-nucleon knock-out  from high energy scattering processes is  the most appropriate venue to probe {\it NN} correlations in nuclei.
Two nucleons can be naturally emitted by two-body mechanisms  \cite{martini}: MEC - two steps interactions probing two nucleons correlated by 
meson exchange currents, and ``Isobar Currents" (IC) - intermediate state $\Delta,~N^*$ excitation of a nucleon in a pair with the pion from resonance decay reabsorbed by the other nucleon. It should be noted that the {\it NN} pairs in these two-body processes may or may not be SRC pairs.\\
One-body interactions can also lead to two-nucleon ejection. This happens when the struck nucleon is in a SRC pair and the high relative momentum in the pair would cause the correlated nucleon to recoil and be ejected as well \cite{benhar2}.\\
 It should also be noted that in both cases final state interactions (FSI) - momenta or charge exchange and inelastic reactions - between the outgoing 
 nucleons and the residual nucleus \cite{giusti} may alter the picture. 

Hadron scattering experiments were extensively performed to probe {\it NN} SRC in nuclei.
In pion-nucleus experiments in the intermediate energy range (incident energy fixed in the $\Delta$-resonance range, 100-500 MeV) the cross section is high and the main contribution is from absorption processes. 
Pion absorption 
is highly suppressed on a single nucleon in the nucleus. Thus, absorption
 requires at least a two-nucleon interaction. The simplest and most frequent absorption mechanism (for A$\ge$12) is on {\it np} pairs (``quasi-deuteron absorption (QDA)": e.g. $\pi^+ + (np)\rightarrow p p$). 
Most of the pion energy is carried away by the ejected nucleons (whose separation energy contributes to the missing energy budget) and part of the momentum can be transferred to the recoil nucleus (missing momentum).
Observation, e.g. from bubble-chamber experiments, of pairs of energetic protons with 3-momentum ${\rm p}_{p1},{\rm p}_{p2}\ge {\rm k}_F$ detected at large opening angles 
in the Lab frame (cos$\gamma \le -0.9$) suggested first hints for SRC in the target nucleus \cite{Bellotti}.

Electron scattering experiments extensively studied SRC.
Experiments of last generation probe SRC by triple coincidence - A($e,e' np{\rm~or~}pp$)A-2 reaction - where the two knock-out nucleons are detected at fixed angles. The SRC pair is typically assumed to be at rest prior to the scattering and the kinematics reconstruction
utilizes pre-defined 4-momentum transfer components determined from the fixed beam energy and the electron scattering angle 
and energy.
{\it NN} SRC are associated with finding a pair of high-momentum nucleons, whose {\it reconstructed} initial momenta are back-to-back and exceed the characteristic Fermi momentum of the parent nucleus, while the residual nucleus is assumed to be left in a highly excited state after the interaction \cite{Arrington}.
 Recent  results from JLab (on $^{12}$C) indicate that $\ge$20\% of the nucleons  (for A$\ge$12) act in correlated pairs. 90\% of such pairs are in the form of high momentum iso-singlet ({\it np})$_{I=0}$ SRC pairs; 5\% are in the form of SRC {\it pp} pairs; and, by isospin symmetry, it is inferred that the remaining 5\% are in the form of SRC {\it nn} pairs \cite{Science}.
 
\begin{figure}
  \vspace{-0.3cm}
   \includegraphics[height=.26\textheight]{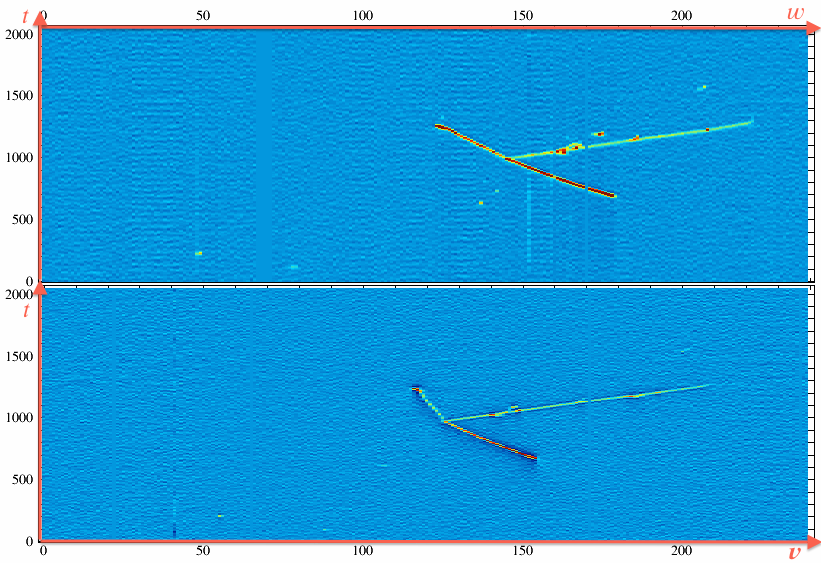}
  \caption{\label{backToback_event_2}{2D views of one of the four ``hammer events", with a forward going muon and a back-to-back proton pair 
  (${\rm p}_{p1}=552~{\rm MeV/c}, {\rm p}_{p2}=500~{\rm MeV/c}$). Transformations from the TPC wire-planes coordinates ({\it w,t} ``Collection plane" [Top], {\it v,t} ``Induction plane" [Bottom]) into Lab coordinates are given in \cite{ArgoNeuT}.}}
\end{figure}
\begin{figure}
  \includegraphics[height=.12\textheight]{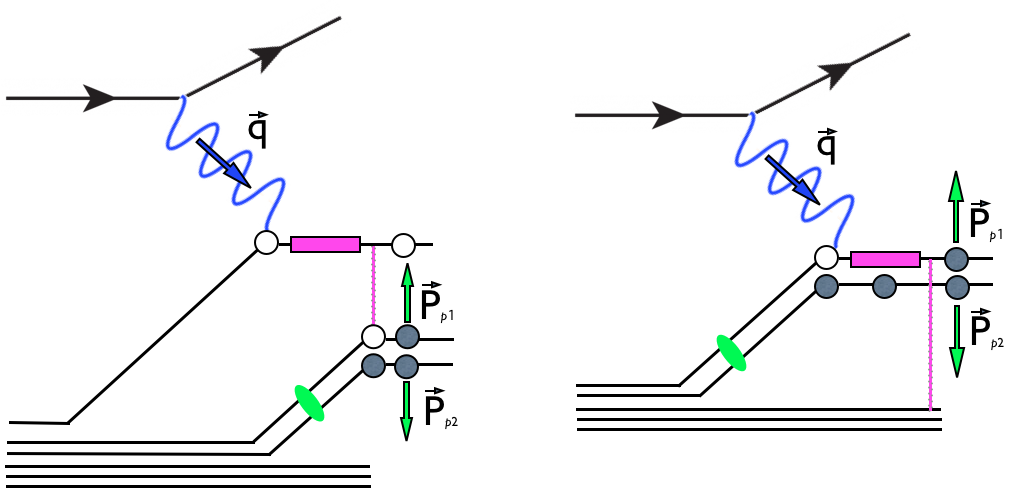}
  \includegraphics[height=.12\textheight]{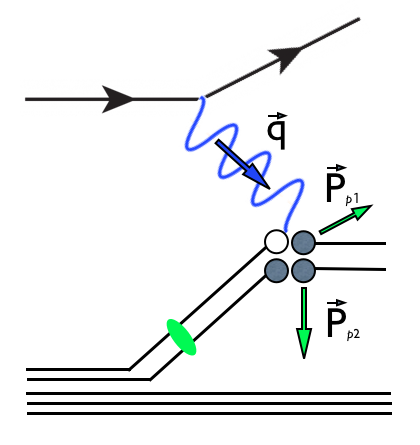}
  \caption{\label{Graphs}{Pictorial diagrams of examples of two-proton knock-out CC reactions involving {\it np} SRC pairs.
   Short range correlated (green symbol) nucleons in the target nucleus are denoted by open-full dots ({\it n-p}), wide solid lines (purple) represent RES nucleonic states, (purple) lines indicate pions. }}
\end{figure}

Neutrino scattering experiments, to our knowledge, have never attempted to directly explore SRC through detection of two nucleon knock-out. 
The main limitation compared to electron scattering comes from the intrinsic uncertainty on the 4-momentum transfer. 
This originates from the {\it a priori} undetermined incident neutrino energy. On the other hand, 
neutrinos can effectively probe the nucleus for its SRC content through both one-body and two-body CC reactions on {\it np} SRC pairs and,
with the advent of LArTPC detectors, two-proton knock-out topologies can be identified unambiguously. The two protons can indeed be detected at any emission angle in the 4$\pi$ sensitive LAr volume and down to energies below the Fermi level (detection threshold in ArgoNeut is $T_p^{\rm thr} = 21$ MeV, i.e. about 200 MeV/c momentum, less than k$_F$ of Ar). \\
To elucidate the role of SRC, we consider here the following neutrino CC interactions leading to two-proton knock-out:\\
\indent - CC RES pionless mechanisms involving a pre-existing SRC {\it np} pair
 in the nucleus. 
For example, (i) via nucleon RES excitation and subsequent two-body absorption of the decay $\pi^+$ by a SRC pair (Fig.\ref{Graphs} [Left]),
 or (ii) from RES formation inside a SRC pair (hit nucleon in the pair) and de-excitation through multi-body collision within the A-2 nuclear system (Fig.\ref{Graphs} [Center]).
Initial state SRC pairs are commonly assumed to be nearly at rest, i.e. $\vec {\rm p}_{p}^{~i}\simeq-\vec {\rm p}_{n}^{~i}$. 
The detection of back-to-back {\it pp} pairs in the Lab frame can be seen as ``snapshots" of the initial pair configuration in the case of RES processes with no or low momentum transfer to the pair.
As noticed, four events in our ($\mu^-+2p$) sample are found with the proton pair in a back-to-back 
configuration in the Lab frame (cos($\gamma$)$<$-0.95, Fig.\ref{cosgamma_vs_mom}). 
Visually the signature of these events gives the appearance of a hammer, 
with the muon forming the handle and the back-to-back protons forming the head. As an example,  
the 2D views from the  two wire planes of the LArTPC for one of these ``hammer" events is reported in Fig.~\ref{backToback_event_2}.
In all four events, both protons in the pair have momentum significantly above the Fermi momentum,
 with one almost exactly balanced by the other, i.e. $\vec {\rm p}_{p1}\simeq-\vec {\rm p}_{p2}$. 
All events show a rather large missing transverse momentum, 
${\rm P}^{\rm T}_{\rm miss}\gtrsim300$ MeV/c. 
These features look compatible with the hypothesis of CC RES pionless reactions involving pre-existing SRC {\it np} pairs.\\
\indent - CC QE one-body neutrino reactions, through virtual charged weak boson exchange on the neutron of a SRC $np$ pair (Fig.\ref{Graphs} [Right]). The high relative momentum will cause the correlated proton to recoil and be ejected. 
 Within impulse approximation, identification of the struck neutron requires a large momentum transfer such that the momentum of the proton emitted in this type of event is much larger than the momentum of the spectator proton in the pair, i.e.:
\begin{equation}
\vec {\rm p}_{p1} = \vec {\rm p}_{n}^{~i} + \vec {\rm q} \gg {\rm k}_F~;~~~~~~~~~~~\vec {\rm p}_{p2} = \vec {\rm p}_{p}^{~i} > {\rm k}_F
\label{mom-transf}
\end{equation}
with both protons exceeding the Fermi momentum, the struck nucleon $p1$ being the higher in momentum 
 and the lower $p2$ identified as the recoil spectator nucleon from within the SRC.
 As mentioned above, momentum transfer in neutrino events is a reconstructed quantity, less precisely determined than in electron scattering experiments.    
 However, with an approach similar to the electron scattering triple coincidence analysis, 
 we determine the initial momentum of the struck neutron from the [Left] equation in 
 (\ref{mom-transf}), i.e. by
 transfer momentum vector subtraction to the higher proton momentum ($\vec {\rm p}_{n}^{~i}=\vec {\rm p}_{p1} - \vec {\rm q})$. 
This procedure is applied to the 
remaining  sub-sample of fifteen ArgoNeuT events ($\mu^-+2p$)
 with both protons above Fermi momentum, after excluding the four hammer events already ascribed to other types of reactions. 
 In most cases the reconstructed initial momentum is found above $k_F$ and 
 with cos$(\gamma^i)<0$ (opening angle $\gamma^i$ between the reconstructed
 struck neutron and the recoil proton in the initial pair), i.e. opposite to the direction of the recoil proton.
\begin{figure}[tbh!]
\vspace{-0.3cm}
  \includegraphics[height=.26\textheight]{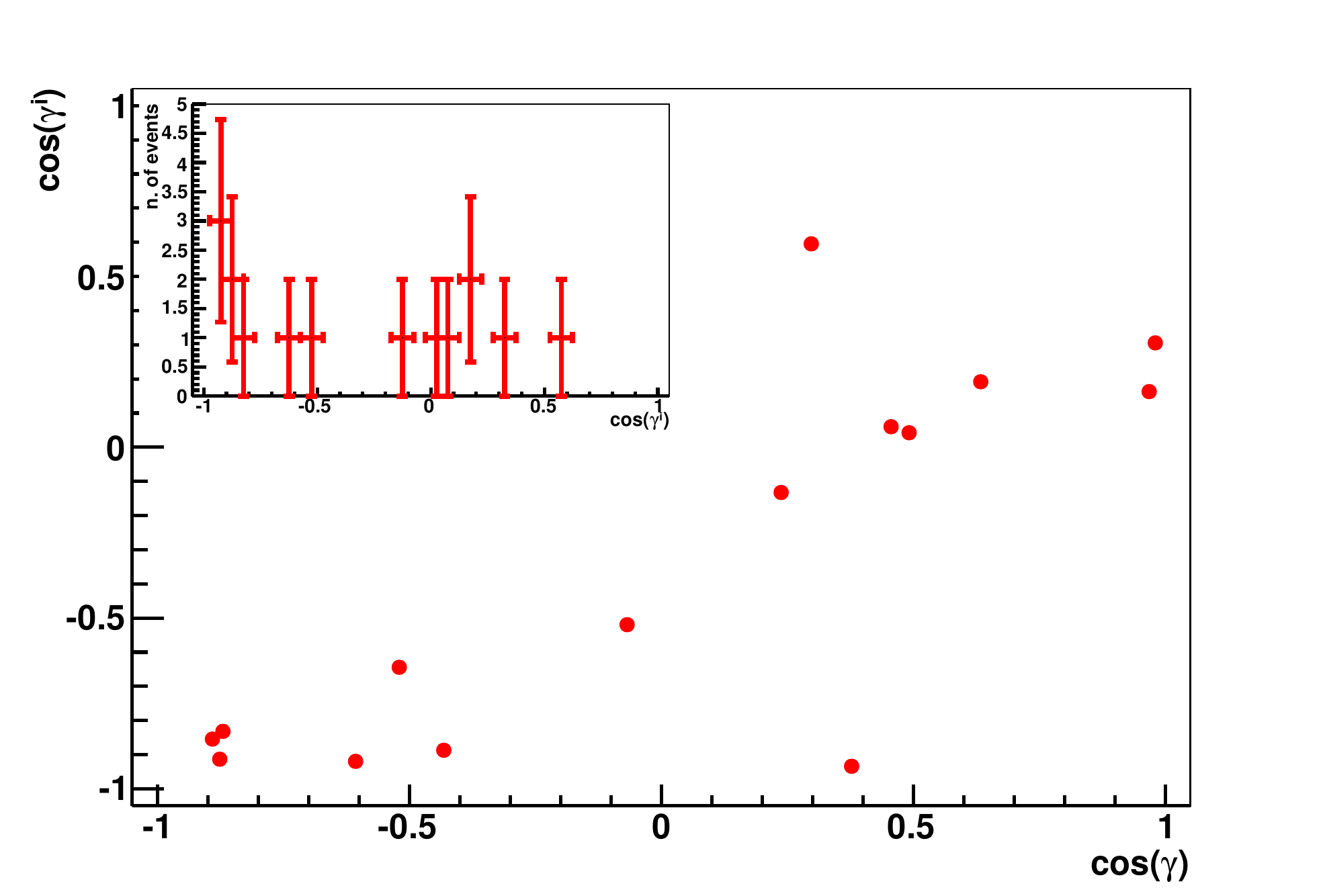}
  \caption{\label{gammai-gamma}{Cosine of the reconstructed opening angle $\gamma^i$ of the initial state  vs.
   cosine of the observed opening Lab angle $\gamma$} of the final state proton pairs (both protons with momentum above k$_F$).
   In the inset, the cos$(\gamma^i)$ distribution of the reconstructed initial pair opening angle.}
\end{figure}
In particular, a fraction of the events exhibit a strong angular correlation peaking at large, back-to-back initial momenta, 
as shown in the inset of Fig.\ref{gammai-gamma}. The bin size includes the effect of the uncertainty in the transfer momentum reconstruction on  the
measurement of cos$(\gamma^i)$.
The measured transverse component of the missing momentum in these events is typically small ($\lesssim250$ MeV/c). 
Under the above assumptions and within the limits of our reconstruction, these events appear compatible as originating 
from SRC pairs through CC QE reactions.\\
 The scatter plot of the cosine of the opening angle $\gamma^i$ in the reconstructed initial {\it np} pair in the nucleus against the cosine of the opening angle 
 $\gamma$ of the {\it pp} pair in the final state observed in the detector is shown in Fig.\ref{gammai-gamma}.
Four of the events mentioned above are those horizontally aligned in the lowest cos$(\gamma^i)\le-0.9$ bin, rather separated from the others.
There is no immediate interpretation for the apparent correlation of the remaining 11 events in the plot. 
Two-step processes such as MEC or IC  involving {\it NN} long-range correlated pair in the nucleus \cite{martini} are obviously active in two-proton knock-out production.
Other mechanisms like interference between the amplitudes involving one- and two-nucleon currents, subject to current theoretical modeling \cite{benhar2}, can also potentially contribute.
 In all cases, protons can undergo FSI inside the residual nuclear system before emerging and propagating in the LAr active detector volume.
In general, however, the emission of energetic, angular correlated proton pairs from FSI appears disfavored.\\
\indent In conclusion, a fraction (four of the thirty events) of the ($\mu^-+2p$) sample detected with ArgoNeuT
are found with the two protons in a strictly back-to-back high momenta (hammer) configuration in the final state. 
Another equivalent fraction is found compatible with a reconstructed back-to-back configuration of a {\it np} pair 
in the initial state inside the nucleus. 
The event statistics from ArgoNeuT is very limited and cannot provide definitive conclusions. 
However, these events suggest that different mechanisms directly involving {\it NN} SRC pairs in the nucleus are active, 
and can be probed efficiently with the LAr TPC technology. 
The inclusion of a realistic and exhaustive treatment of SRC in the one- and two-body component of the nuclear current is 
a great challenge in current theoretical modeling and subsequent MC implementation. 
More accurate and detailed MonteCarlo neutrino generators are deemed necessary for comparisons with LAr data. 
Future larger mass and high statistics LAr-TPC detectors have the opportunity to clarify the issue.
We hope the ArgoNeuT data will encourage more studies in this area.\\
\indent We gratefully acknowledge the cooperation of the MINOS Collaboration in providing their data for use in this analysis. 
We acknowledge the support of Fermilab, the Department of Energy, and the National Science Foundation in ArgoNeuTÕs construction, operation, and data analysis.


\end{document}